\documentclass{article}

\usepackage{arxiv}

\usepackage[utf8]{inputenc} 
\usepackage[T1]{fontenc}    
\usepackage{hyperref}       
\usepackage{url}            
\usepackage{booktabs}       
\usepackage{amsfonts}       
\usepackage{nicefrac}       
\usepackage{microtype}      
\usepackage{graphicx}
\usepackage{acro}
\usepackage{array}
\usepackage{amsmath} 

\author{Peter Horoschenkoff$^{1,2}$ \and Jasper Rödiger$^2$ \and  Martin Wilske$^2$}

\newcolumntype{C}[1]{>{\centering\let\newline\\\arraybackslash\hspace{0pt}}m{#1}}
\newcolumntype{L}[1]{>{\raggedright\let\newline\\\arraybackslash\hspace{0pt}}m{#1}}

\title{A new control- and management architecture for SDN-enabled quantum key distribution networks}

\DeclareAcronym{qkd}{
  short=QKD,
  long=quantum key distribution,
}

\DeclareAcronym{qkdn}{
  short=QKDN,
  long=quantum key distribution network,
}

\DeclareAcronym{kms}{
  short=KMS,
  long=key management system,
}

\DeclareAcronym{pqc}{
  short=PQC,
  long=Post-Quantum Cryptography,
}

\DeclareAcronym{sdn}{
    short=SDN,
    long=software-defined networking,
}

\DeclareAcronym{cm}{
    short=CM,
    long= control- and management,
}

\DeclareAcronym{km}{
    short=KM,
    long=key management,
}

\DeclareAcronym{psk}{
    short=PSK,
    long=pre-shared-keys,
}

\DeclareAcronym{cms}{
    short=CMS,
    long=control and management as a Service,
}

\DeclareAcronym{sp}{
    short=SP,
    long=separately-protected,
}

\DeclareAcronym{dos}{
    short=DoS,
    long=Denial-of-Service,
}
 
\begin{document}

\maketitle

\section*{Abstract}
This paper aims to address the challenge of designing secure and high performance Quantum Key Distribution Networks (QKDN), which are essential for encrypted communication in the era of quantum computing. Focusing on the control and management (CM) layer essential for monitoring and routing, the study emphasizes centrally managed software defined networks (SDN).
We begin by analyzing QKDN routing characteristics needed for evaluating two existed architectures and the proposed, new CM layer implementation. Following the theoretical analysis, we conduct a discrete-event based simulation in which the proposed architecture is compared to an existent serving as performance-baseline. The results provide recommendations based on use cases for which different architectures show superiority and offer valuable insights into the development and evaluation of CM architectures for QKDNs.

\section{Introduction}
Quantum-safe technologies are needed for enabling encrypted communication in the age of quantum computing \cite{Mehic}. As a promising candidate, \ac{qkd} is becoming increasingly important as it is immune to any computational threat. As of today, \ac{qkd} enables key distributions between two points which can typically be 100 km apart due to transmission losses \cite{Mehic}. To enable the usage of QKD by multiple users over arbitrary distances, multiple QKD links need to be combined to a \ac{qkdn}. It is useful to structure a \ac{qkdn} into four different layers in a hierarchical alignment: the quantum layer, the \ac{km} layer, the application layer and the \ac{cm} layer \cite{ITU3803}. 

Each layer is assigned a different task. In the lowest, the quantum layer, pairs of QKD devices, a sender and a receiver, are connected by a QKD link generating symmetric random bit strings according to a given (QKD) protocol. The task of the \ac{km} layer above is to generate keys from the received bit sequence, to store and forward them. The relay (or transport) of keys is characterized by a key distribution scheme and enables key sharing on a wide network topology using so-called trusted nodes \cite{ITU3803}. This task is executed by the \ac{kms} existent in every node. Cryptographic applications using secure keys, to en- and decrypt user traffic, are located in the application layer. They request secure keys from the underlying \ac{km} layer. The \ac{cm} layer is monitoring the network in terms of the node- and link status. Further, it incorporates routing capabilities, including the deployment of a routing algorithm to determine the most optimal path for key transport within the \ac{km} layer. Furthermore, it can be extended to support additional tasks commonly required in carrier networks, such as accounting and authentication capabilities. 

While the interaction of the \ac{cm}- with the \ac{km} layer, i.e. \textbf{what} information is exchanged defined by routing algorithms, is the subject of much research \cite{wdm_routing, Sharma, stoachstic_routing, ada_china_routing, qos_routing}, the implementation of the \ac{cm} layer (its architecture) is still quite unexplored, i.e. \textbf{how} information is transported to and from the \ac{cm} components. As these networks are to be used in highly secure network scenarios, it is crucial to cover this layer in the same thorough manner. The goal of this work is to contribute in this gap by investigating different implementations of the \ac{cm} layer in terms of their security and performance. The focus of this work is on centralized managed \ac{qkdn}.

Centralized managed networks can be efficiently enabled by \ac{sdn}. The core idea of \ac{sdn} is the disassociation of the control- and data plane. The control plane determines the process of handling and forwarding data traffic and the data plane follows the instructions provided. An interface between both is provided by their management layer that is used to configure the data plane. The control plane can be, without considering redundancy, represented by a single instance while the data plane is located in every network node. Additionally, each node has an \ac{sdn}-Agent that manages its unique data plane, receiving instructions from the \ac{sdn}-Controller. This approach comes with multiple advantages: highly efficient use of resources, a centralized management enabling the straight forward integration of different applications as well as a global view on the network \cite{sdn_survey}.

This trend, towards the development of such centralized managed networks, has also reached the \ac{qkdn} sector \cite{sdn_qkd_trend1, sdn_qkd_trend2, ITU3805, etsi_015, madrid_qkdn, sdn_uk}. Initial studies focused on extending existent protocols used in \ac{sdn} with \ac{qkd} generated keys \cite{sdn_qkd_trend1, sdn_qkd_trend2} and were sustained, by standardization works on the requirements and general operating procedures \cite{ITU3805} as well as abstraction models and workflows for \ac{sdn}-enabled \ac{qkdn} \cite{etsi_015}. Nevertheless, the optimal architecture, i.e. the implementation of the \ac{cm} layer, for interfacing with the different layers while deploying the \ac{sdn} functionality remains a field of optimization.  

In the Madrid \ac{qkdn} \cite{madrid_qkdn}, a QKD \ac{sdn} controller (QSDN-Controller) is used centrally on a dedicated “controller node”. Each network node is connected to it on a point-to-point basis. An \ac{sdn}-Agent assigned to each node manages the directives from the QSDN-Controller. Without the controller being attached to the quantum layer, it is not possible to establish a \ac{qkd}-secured communication for the traffic assigned to it. However, it is possible to establish a secure connection by means of \ac{psk} or Post-Quantum Cryptography (PQC) algorithms. Although establishing a QKD secured \ac{cm} layer is beyond the scope of \cite{madrid_qkdn}, it demonstrates the significant potential of \ac{sdn}-enabled \ac{qkdn}. Therefore, it perfectly serves as performance baseline for a comparative evaluation of other \ac{cm} architectures, as it does not use the QKD technology to secure the \ac{cm} traffic. Hence, in this work, this architecture is referred to as \textit{\ac{sp} architecture} representing the most ideal performance a \ac{cm} layer implementation can achieve.

A different architectural approach is chosen in \cite{sdn_uk}. Again, a centralized (\ac{sdn}) controller holds global knowledge of the network. In difference to the previous implementation the controller node additionally features a QKD-Module and a \ac{kms}. To establish a secure communication to the \ac{cm} layer, the \ac{sdn}-Agents and the QSDN-Controller are attached as SAEs (Secure Application Entities) to the respective \ac{kms}. An SAE is a generic description for an application requesting and receiving keys from the \ac{km} layer \cite{etsi_014}. The \ac{sdn}-Agent can now establish an encrypted connection to the QSDN-Controller via the application layer using a previously distributed key from the \ac{km} layer. As the \ac{cm} layer is using the key provisioning service of the \ac{qkdn}, we refer in this work to the architecture as \textit{Control-and-Management as-a-Service (CMS) architecture}. As \ac{qkdn} are foreseen to be deployed in high secure scenarios, preventing metadata leakage is of high priority, since it can potentially provide an adversary with actionable intelligence that could be exploited to launch targeted attacks or to gather sensitive information about user behavior. A security problem arises when an eavesdropper is located in a position to intercept communication between devices and the \ac{sdn} controller (e.g. directly in-front of the central connected \ac{sdn} controller), thereby gaining access to sensitive metadata. However, despite an encryption of the payload, the eavesdropper can still deduce valuable information, including the number of devices connected to the network and the frequency of interactions between individual devices and the controller, which can be used to compromise the security of the network.
In Chapter \ref{chapter:routing_in_qkdn}, we provide a more detailed analysis of both architectures in terms of their security and performance features.

In this paper, we present a novel, secure way to implement the \ac{cm} layer in a centralized managed \ac{qkdn} that is protecting the KM layer from exposing metadata, and efficiently authenticates the \ac{cm} traffic. By comparing the new approach to two existing architectures, we demonstrate the ad- and disadvantages through a theoretical analysis. We further demonstrate its feasibility and performance through a simulation study, by benchmarking the proposed against the \ac{sp} architecture. Summarized, in this work we show:

\begin{itemize}

    \item A comparison of different implementations of the \ac{cm} layer in view of previously identified characteristics
        
    \item A new implementation of the \ac{cm} layer which aims to maximize the secure communication with this layer
    
    \item A simulation evaluating two different \ac{cm} layer architectures at various key generation rates
    
    \item Recommendations, based on use cases in which different architectures show advantages.
    
\end{itemize}

The paper is organized as follows: Section \ref{chapter:towards_routing_in_qkdn} describes routing characteristics of \ac{qkdn}. Section \ref{chapter:routing_in_qkdn} presents a new architecture as well as a comparison of different architectures for the \ac{cm} layer. The simulation and its results are discussed in Section \ref{chapter:simulation}. The work is summarized and concluded in Section \ref{chapter:summary}.

\section{Differences between QKD- and classical telecommunication networks}\label{chapter:towards_routing_in_qkdn}

For the design and evaluation of \ac{cm} architectures in trusted-node \ac{qkdn}, we begin by analyzing the routing characteristics that need to be considered. 
We build upon the listed characteristics in \cite{ada_china_routing} with our own and further extended the work by drawing new conclusions from these.

Routing in \ac{qkdn} greatly differs from routing in classical telecommunication networks (CTN). A summary of the characteristics is given in Table \ref{tab:routing_characteristics_qkdn}. Based on these characteristics we draw the following conclusions.
The increased demands on packet processing must be met with sufficient resources, which can be achieved by using suitable, high-performance hardware or parallel processing techniques. 
The performance of a KMS not only depends on the efficient encryption and decryption of messages, but also on the availability of sufficient keys. Keys according to link-specific requirements must be available \cite{ada_china_routing}.

As the successful routing depends on the key generation of multiple nodes, within an optimal path, the combined bandwidths of quantum- and classical channels along that path need to be considered. Congestion avoidance therefore begins in the network design and needs to encompass all involved layers. This involves strategically positioning the nodes to account for their high sensitivity to distance in dependence of their key consumption, as well as implementing path diversity to ensure robustness in the event of a failure in view of specific security requirements. 

The traffic dependencies between the application and \ac{km} layers can lead to congestion in both layers, either through excessive key requests or delayed key delivery. To mitigate this, the dependency must be made unidirectional, allowing the \ac{km} layer to cause congestion in the application layer, but preventing the application layer from impacting the \ac{km} layer. A single node in the application layer should not be able to induce congestion on nodes in the \ac{km} layer. To achieve this, mechanisms must be in place to prevent application layer congestion from affecting the \ac{km} layer, making traffic engineering a crucial aspect. This also includes routing protocols that facilitate appropriate load balancing within the \ac{km} layer.

Based on the distinctive characteristics of a \ac{qkdn}, it can be inferred that the routing protocols and network implementations utilized in CTN are not directly applicable to \ac{qkdn} due to the substantial differences between the two network types. Furthermore, the \ac{km} layer is more vulnerable to attacks due to its increased complexity and its critical role in securely relaying keys with high performance and precision, making it a more attractive target compared to the application layer. Consequently, effective countermeasures must be implemented to address the increased vulnerability.

\begin{table*}[htp]
    \centering
    \begin{tabular}{|c|L{5.8cm}|L{5.8cm}|}
        \hline
         Name & Classical Telecommunication Networks & Quantum Key Distribution Networks  \\
        \hline
         \textbf{Packet Processing} \cite{ada_china_routing} & Packets are processed according to a locally stored routing table or received routing vector to determine the receiver of the next hop & 
        The routing process involves the forwarding of keys, whereby trusted nodes send and receive encrypted packets selecting the keys in a way such that the overall path between two requesting users is optimized \\
         \hline
         \textbf{Forwarding Capacity} \cite{ada_china_routing} & The network bandwidth influences the ability to forward packets & Forwarding capacity is limited by the minimum of either the key generation rate and the residual key volume, or the network bandwidth \\
         \hline
         \textbf{Routing success rate} \cite{ada_china_routing} & Exceeding node bandwidth and processing capacity causes network congestion & 
         Network congestion occurs when either the bandwidth of the quantum channel or the classical channel is exceeded \\
        \hline
        \textbf{Traffic dependency} &  Traffic can be assumed independent & Traffic in the \ac{km} layer is dependent on the amount of traffic in the application layer \\
        \hline
    \end{tabular}
    \caption{Comparison of the routing characteristics of Quantum Key Distribution Networks and Classical Telecommunication Networks.}
    \label{tab:routing_characteristics_qkdn}
\end{table*}

\section{Control- and Management architectures}\label{chapter:routing_in_qkdn}

In view of the above identified characteristics and conclusions of \ac{qkdn}, we investigate different implementations of the \ac{cm} layer, i.e. \textbf{how} information is transported to and from the \ac{cm} layer. 

\subsection{SDN-enabled QKD networks}
As \ac{qkdn} can be efficiently enabled by the \ac{sdn} technology, a central instance needs to collect and manage information on the network to determine best routing choices, i.e. the QSDN-Controller \cite{madrid_qkdn, sdn_uk}. The QSDN-Controller is an \ac{sdn} controller that belongs to the \ac{cm} layer within the \ac{qkdn} architecture and is located in a controller node. It exchanges \ac{cm} information with the \ac{sdn}-Agent. The \ac{sdn}-Agent is embedded in the QKD \ac{sdn} node (QSDN node) and collects or distributes the information from the other network layers. The QSDN node architecture further includes at least one QKD-Module in the quantum layer, a KMS in the \ac{km} layer, potentially a network encryptor (NE) or other SAE in the application layer. Figure \ref{fig:overview_routing_architectures} depicts three different implementations of the \ac{cm} layer, i.e. how the controller node receives the \ac{cm} information from the QSDN nodes and vice versa. Other tasks for establishing a secure key between two distant users using the QKD technology, e.g. the KMS or QKD-Modules, are not shown for sake of simplicity. The following analysis of these three \ac{cm} architectures concentrates on evaluating their security and performance characteristics. The security aspects are focused to metadata leakage on the network's topology, such as the amount of nodes or infer node activities/priorities by analyzing the frequency of requests from the reactive routing protocol, as well as authentication and \ac{dos} vulnerabilities.

\subsection{Separately-protected Architecture}
In Figure \ref{fig:overview_routing_architectures}, the Separately-protected (\ac{sp}) architecture is depicted on the left, illustrating an implementation of the \ac{cm} layer that utilizes alternative security technologies, rather than QKD, to secure the \ac{cm} traffic. In this architecture, every node instance has a dedicated connection to the QSDN-Controller, i.e. physical point-to-point connection. The QSDN-Controller, which is solely located in the controller node, exchanges information with the \ac{sdn}-Agent existent in every network node. In this architecture, the controller node does not attribute a connection to the quantum layer, thereby precluding \ac{qkd}-secured communication. However, it would be possible to establish an encrypted connection by means of \ac{psk} or PQC-algorithms. Therefore, this architecture represents the most ideal performance a \ac{cm} layer implementation can achieve, as it does not use the QKD technology to secure this traffic which may lead to earlier network congestion due to an increased key consumption.

This architecture offers a significant advantage due to its direct correspondence between the physical and logical architecture. This design is particularly beneficial for implementing routing algorithms that rely on flooding the network with status information, as it either prevents (for every node having an own channel) or reduces (for multiple nodes sharing one channel) this network-flooding. Furthermore, despite the potential challenges of implementing multiple dedicated channels in large-scale networks, this approach provides a highly redundant network topology, ensuring that the QSDN-Controller remains accessible even in the event of a channel failure, thanks to adjacent nodes relaying messages through alternative paths.

In this architecture, a potential eavesdropper positioned in front of the QSDN-Controller may intercept information from the packet headers, as each node maintains a direct connection to the controller. This enables the eavesdropper to collect metadata on the network's topology. Furthermore, the QSDN-Controller is highly exposed and constitutes a single point of failure, making it an favourable target for potential attacks. Proper countermeasures must be put in place.
Since the \ac{cm} traffic also contains routing commands, it needs to be at least authenticated. Respective certificates need to be installed and managed in all nodes during the networks lifetime. The controller node, however, may face difficulties in achieving this due to the absence of an interface to the quantum layer.

Relaying data through alternative network interfaces, such as network encryptors or connected SAEs, to the QSDN-Controller, rather than using the network interface of the \ac{sdn}-Agent itself, necessitates QSDN nodes to interface with the \ac{qkdn} layer, thereby increasing the layer's utilization. This also includes an increased processing pressure of the devices forwarding the traffic.  
Another option is to utilize a separate and independent network infrastructure (perhaps an already existent one) that operates in isolation from the \ac{qkdn} network, such as a 5G network or another carrier network. This approach offers optimal performance, as the \ac{cm} traffic is handled separately. From a security perspective, it also provides the advantage of being decoupled from the redundancy attribute of the application layer. Nevertheless, again another network infrastructure and -interface is needed, hereby increasing the attack surface of the \ac{qkdn}.

\subsection{Control-and-Management-as-a-Service Architecture}
The architecture shown in the middle of Figure \ref{fig:overview_routing_architectures} embodies the principle of the architecture proposed in \cite{sdn_uk}, i.e. the \ac{cms} architecture.  Again, every node holds a dedicated connection to the QSDN-Controller. Compared to the \ac{sp} architecture, the main difference is that the QSDN controller attributes a connection to the quantum layer. This enables the nodes to establish a secure channel via the \ac{km} layer to the controller node, as the \ac{sdn}-Agents and the QSDN-Controller are attached as SAE to the respective KMS. Further, this enables an easier authentication of the \ac{cm} traffic throughout the networks lifetime.
Nevertheless, it leaks the same amount of metadata, as the \ac{sp} architecture and also features a highly exposed network controller that constitutes a single point of failure. Again, appropriate countermeasures must be implemented to mitigate potential risks. As this architecture uses two different network parts of the \ac{qkdn}, namely the \ac{km} network and the application layer, to establish the connection to the controller node, it relies on the redundancy of both parts in the event of a \ac{dos} attack. Depending on the topology of the network handling the \ac{cm} traffic this approach may (as above) be effective for routing protocols that rely on network flooding.

Securing the \ac{cm} traffic with \ac{qkd} keys comes with an increased processing pressure in the \ac{km} layer as well as with adjustments on the link specific a number of available keys. Dependent on the packet-to-key-ratio, i.e. how often a key is changed in the packet encryption process, the adjustments may also influence the initial design of the QKDN with its components. As in the \ac{sp} architecture, the \ac{cm} traffic from the \ac{sdn}-Agent to the QSDN-Controller can be relayed via an alternative network infrastructure, which (as above) leads to both: an increased attack surface due to the additional network interface and an improved resilience against \ac{dos} attacks due to the independent infrastructure.

\subsection{Control-and-Management-via-KMS architecture}
\begin{figure*}[htp] 
    \centering
\includegraphics[width=\textwidth]{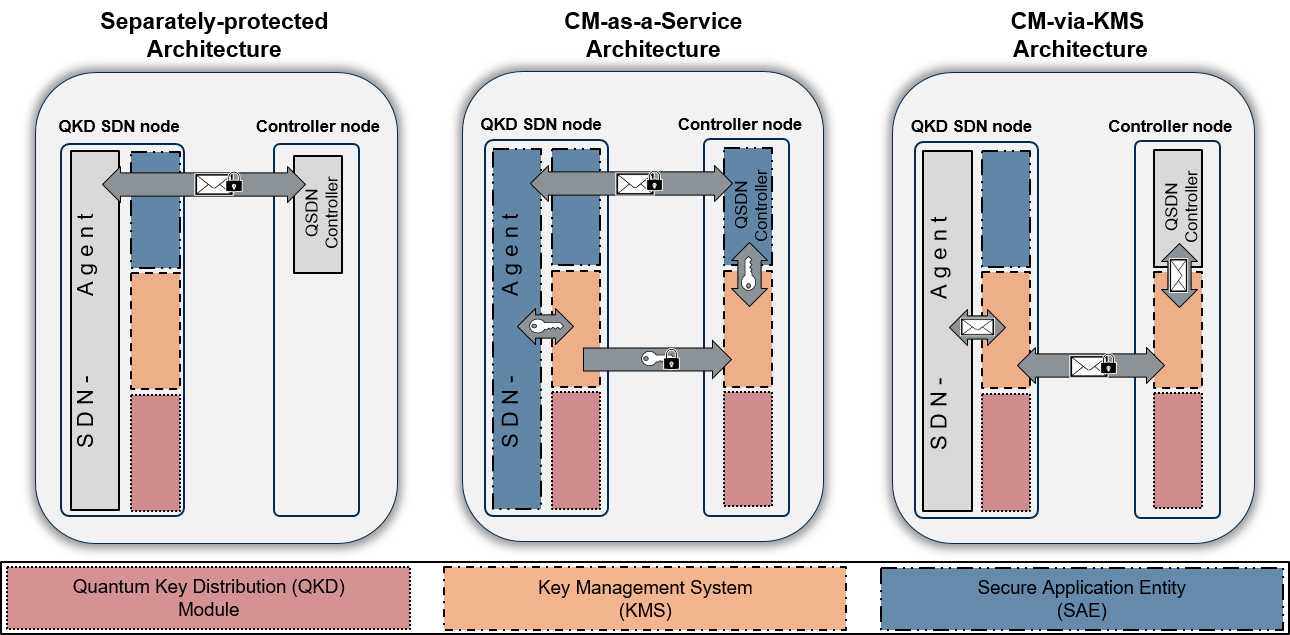}
    \caption{This Figure shows different implementations of the Control- and Management (CM) layer for \ac{qkdn}. Other tasks of the network devices are not depicted. To implement the \ac{sdn} functionality, a centralized QKD \ac{sdn}-Controller (QSDN-Controller) hosts global knowledge on the network and its devices. It is linked to the \ac{sdn}-Agent residing in each node, thereby establishing a unified control plane that enables efficient network management and orchestration. 
    On the left side, every node has a dedicated connection to the QSDN-Controller it uses crypto primitives beside QKD to secure its \ac{cm} traffic, i.e. the \ac{sp} architecture \cite{madrid_qkdn}. In the middle, the \ac{sdn}-Agent as well as the QSDN-Controller are attached as SAEs such that a secure connection via the \ac{km} layer can be established which is deployed in another layer, i.e. the \ac{cms} architecture \cite{sdn_uk}. On the far right, the \ac{cm} traffic is relayed via the \ac{km} layer in the same manner of a key transport towards the QSDN-Controller, i.e. the \ac{cm}-via-KMS architecture. Here, the KMS holds the capabilities of key forwarding and \ac{cm} traffic relaying.}
    \label{fig:overview_routing_architectures}
\end{figure*}

The new architecture that is proposed in this work is shown in Figure \ref{fig:overview_routing_architectures} on the right, i.e. the \textit{\ac{cm}-via-KMS architecture}. Again, a secure channel is established but without setting the QSDN-Controller and \ac{sdn}-Agents as SAEs. While CM messages in the \ac{cms} architecture are relayed via an alternative management network, messages in this architecture are relayed in the \ac{km} layer. More specifically, the messages exchanged with the QSDN-Controller are disseminated throughout the network in a manner analogous to key transports resulting from key requests. Notably, the KMS serves a dual purpose, not only facilitating key transport functions but also forwarding \ac{cm} traffic received from an internal interface, thereby integrating \ac{km}- and \ac{cm} traffic forwarding capabilities within a single entity. The interface for receiving the \ac{cm} traffic could be the same as the existent one used for receiving the random-number-generated key, which is used to secure the user traffic.

This approach comes with multiple advantages in terms of security for the \ac{km} layer. Firstly, no longer the previous discussed metadata is revealed. An eavesdropper located directly in front of the QSDN-Controller now gains no more information about the metadata than between any other connection in the \ac{km} layer, as the traffic is single-hop and point-to-point. Secondly, no additional interface to the application-layer or to the dedicated management network is needed. This approach thus does not result in an expanded attack surface. Further, the use of the \ac{km} layer for exchanging \ac{cm} messages reduces the authentication demands. As a result, when the \ac{km} layer provides authentication, the \ac{cm} layer's authentication is given by design, as \ac{cm} messages are only relayed between already authenticated/trusted nodes. A last advantage is that sophisticated attacks, such as node-specific \ac{dos} attacks are no longer possible, e.g. blocking only the traffic of node A for deflecting another (physical) attack on node B. These security features come at the cost of an increased processing pressure on the nodes located in the \ac{km} layer. This is particularly relevant for the gateway node, which is the node, i.e KMS, connecting the controller node to the rest of the \ac{qkdn}. As the management traffic of every network instance needs to relayed by this component, the additional load on the gateway node and the overall \ac{km} layer needs to be investigated, cf. Chapter \ref{chapter:simulation}. Similar to the \ac{cms} architecture, securing the \ac{cm} traffic with \ac{qkd} keys increases the processing pressure in the \ac{km} layer and the overall key consumption in the \ac{qkdn}. This processing pressure can be reduced using higher packet-to-key ratios when deploying cryptographic primitives beside One-Time-Pad (OTP).
Since the \ac{cm} traffic is relayed via the \ac{km} layer, the robustness against \ac{dos} attacks is contingent upon the \ac{km} layer's ability to withstand such attacks. This may show a disadvantage. On the other side, an available communication with the QSDN-Controller may be useless if the communication with other \ac{km} nodes fails.

This architecture shields the \ac{km} layer from metadata exposure as well as by authenticating the \ac{cm} traffic by its design. This is particularly crucial when the network topologies of the \ac{km} layer and the application layer diverge. Further, this protection is essential, as the \ac{km} layer is inherently more vulnerable to attacks compared to the application layer, cf. Section \ref{chapter:towards_routing_in_qkdn}.

Summarized, we distinct between three different implementations of the \ac{cm} layer. Whereas in the \ac{sp} architecture the missing quantum layer causes authentication and encryption difficulties, the \ac{cms} architecture establishes a secure channel to the QSDN-Controller by encrypting it with QKD keys. The new proposed architecture relays \ac{cm} messages in the same manner as a key transport to enhance the security for the \ac{km} layer and thus for the network.  

\section{Simulation}\label{chapter:simulation}
In order to evaluate the performance of the new, proposed architecture different simulations are carried out and analyzed. The simulations (scenarios A-D, as outlined in Figure \ref{fig:sim_results}) each comprise a routing protocol and architecture, and are performed for a range of key generation rates. The other layers remain the same. Before we can compare the different implementations, we examine various routing protocols that are then used in the different network architectures.

\subsection{Selected routing protocols and architectures}

Routing protocols can be classified as static (non-adaptive) or dynamic (adaptive). Static protocols use fixed rules and do not adapt to network changes, while dynamic protocols adjust routes based on real-time network information. Due to similarities with mobile ad hoc networks (MANETs) \cite{Mehic}, research on \ac{qkdn}s has focused on dynamic protocols. Dynamic protocols are either reactive or proactive. Reactive protocols determine routes on-demand, with each packet requesting its own routing vector. Proactive protocols periodically update the network with routing information. In source routing, the sender determines the entire path, whereas in distributed routing, intermediate nodes incrementally determine the next optimal destination \cite{survey_routing_protocols_2}.

Our primary objective is to demonstrate the feasibility of the proposed architecture, specifically its ability to handle the increased processing pressure as already inherent in \ac{qkdn}, as discussed in Chapter \ref{chapter:towards_routing_in_qkdn}. We aim to determine whether the security benefits justify the costs. To this end, we only need to compare the performance of the \ac{cm}-via-KMS architecture against the performance of the \ac{sp} architecture. First, as the \ac{sp} architecture achieves optimal performance by securing \ac{cm} traffic without relying on QKD keys and second, as a simulation of the \ac{cms} architecture would not provide new insights, as it is already encompassed by both the \ac{sp} and the \ac{cm}-via-KMS architecture. This is explained in the following: depending on the ratio of encrypted packets per key (deploying AES-encryption), the performance difference between the \ac{sp}- and \ac{cms} architecture can be negligible. When using the identical packet-to-key ratio for the \ac{cm} traffic as for the user traffic, the amount of keys required for securing the \ac{cm} traffic is negligible compared to key material needed for the application layer. When using a $1:1$ packet-to-key ratio, the \ac{cms} architecture becomes identical to the \ac{cm}-via-KMS architecture approach, as the same amount of keys is required for requests to and from the controller, in addition to the use of an external management network. Given the negligible impact of additional \ac{cm} traffic, simulating the \ac{cms} architecture would not provide new insights, whereas simulating the \ac{sp} architecture offers a true best performance benchmark, independent of the scenario dependent packet-to-key ratio.

The routing protocols we selected are a source-reactive and a distributed-proactive routing protocol, which are used in either the \ac{sp}- or the proposed architecture. The exact simulation framework, i.e. the used assumptions and parameters, can be found in the Appendix \ref{chapter:appendix_sim_body}.
As a full description of a QKDN would include a large number of parameters we have limited our characterization of a QKDN to four different core parameters that allow us to compare the simulation w.r.t. our previous discussed characteristics. These are: $\overline{T_{\text{msg}}^{\text{ne}}}$ describes the average message latency of the NE message, $\overline{T_{\text{key}}}$ describes the average time between a key request and the key arrival,  $\overline{T_{\text{msg}}^{\text{km}}}$ describes the average message latency of the \ac{km} messages and $\overline{N_{\text{msg}}^{\text{km}}}$ describes the average queue length of \ac{km} messages.

\subsection{Simulation Results}
\begin{figure}[ht]
    \centering
    \includegraphics[width=0.5\textwidth]{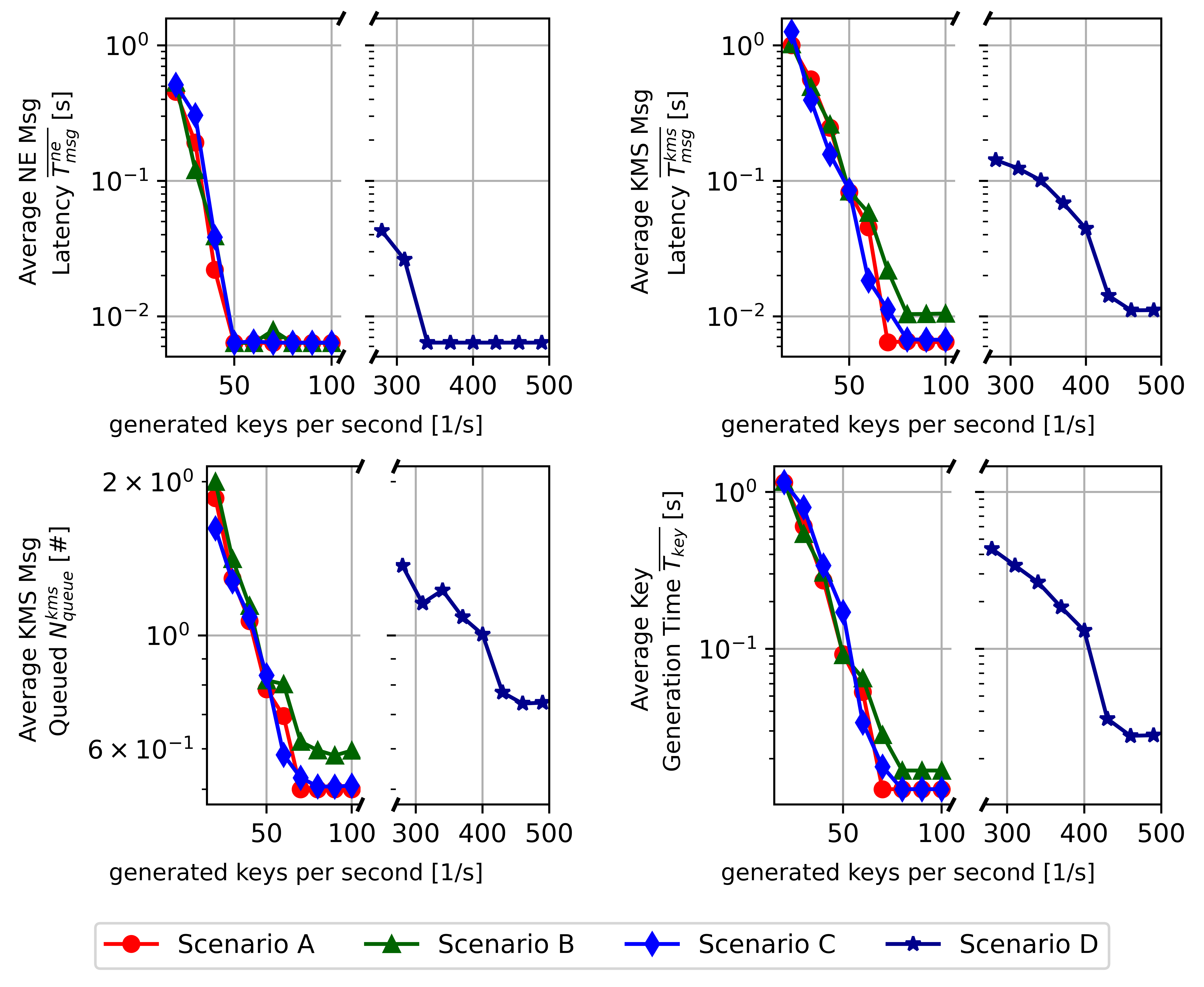}
    \caption{We examined the impact of varying key generation rates on the performance of different \ac{qkdn}s, each deploying a combination of routing protocol and \ac{cm} architecture. Scenarios A and B use a distributed proactive routing protocol and a source-reactive routing protocol in the \ac{sp} architecture. Scenarios C and D use the \ac{cm}-via-KMS architecture with a distributed proactive routing protocol and a source-reactive routing protocol respectively.}
    \label{fig:sim_results}
\end{figure}

Figure \ref{fig:sim_results} presents the measured parameters for network scenarios A-D, plotted against various key generation rates (expressed in keys per second, kps).
The \ac{sp} architecture is utilized in scenarios A and B, where scenario A is characterized by a distributed-proactive routing protocol and scenario B by a source-reactive routing protocol. In contrast, scenarios C and D employ the proposed architecture, with scenario C implementing a source-reactive routing protocol and scenario D implementing a distributed-proactive routing protocol.
Exact predictions of the simulator are obtained by running the simulation multiple times with different random seeds and calculating the average value (c.f. Chapter \ref{chapter:sim_framework}). The main objective of this simulation study is to demonstrate the feasibility of the proposed \ac{cm} architecture, with a focus on obtaining a first-order estimate of the selected parameters rather than an exact average result of a single specific \ac{qkdn} architecture. However, despite the limitation of only running the simulation with only one seed, the results still offer interesting and valuable insights, serving its intended purpose. 

A comparison of the $\overline{T_{\text{msg}}^{\text{ne}}}$ parameter shows the value for the generated-keys-per-second, where users will first perceive a performance degradation. This specific value is referred to as cut-off point. For scenarios A, B and C this is achieved at less than $50$ kps and for scenario D at less than $340$ kps. At lower values, the user may experience noticeable delays or even a complete loss of encryption functionality. The different values of the cut-off points indicate a strong dependency on the key consumption of the routing protocol used for the \ac{cm}-via-KMS architecture, which can be observed for each of the selected parameters. The processing pressure is quantified by the $\overline{N_{\text{msg}}^{\text{km}}}$ parameter. Increased processing pressure is observed in scenarios B and D, where additional routing vector requests cause an increased processing requirement. In line with expectations, scenario D shows the highest processing pressure. In scenario C, on the other hand, the impact of \ac{cm} traffic is minimal compared to the \ac{km} traffic, why only very slight differences are noticed. The earlier cut-off points compared to the $\overline{T_{\text{msg}}^{\text{ne}}}$ parameter indicates the traffic dependency between both layers.
Notably, the performance ranking of the architectures remains consistent, as long as there are sufficient keys available, i.e. above the cut-off points. This can be investigated by the parameters describing the \ac{km} layer ($\overline{N_{\text{msg}}^{\text{km}}}$, $\overline{T_{\text{msg}}^{\text{km}}}$ and $\overline{T_{\text{key}}}$), as the curve order remains static. The performance ranges from the lowest in scenario D, followed by scenario B, and then the top-performing scenarios C and A. Because the parameters are calculated as the average of each node's individual parameter, an order difference is observed as soon as the keys become a scarce resource, i.e. below the cut-off points. First, as oscillating curves are difficult to capture and second, as the measured average value is seed specific.
In terms of the latency in the \ac{km} layer ($\overline{T_{\text{msg}}^{\text{km}}}$), the distributed-proactive routing protocols (scenario B and D) perform better than their counterpart, the source-reactive routing protocols (scenario A and C). This is a consequence of our previous assumption, where the routing vector remains static and does not adapt to the current network status, resulting in a lookup time for determining the next receiver in the routing table that is shorter than the round-trip propagation time to and from the QSDN-Controller for requesting the routing vector. In other words, the communication with the QSDN-Controller in the source-reactive protocols only delays the arrival times of the keys. As the key can only be used when the receiving node confirmed the key relay, the values for $\overline{T_{\text{key}}}$ in the scenarios B and D are roughly twice the value of $\overline{T_{\text{msg}}^{\text{km}}}$ which can be attributed to the propagation delay of the acknowledgement back to the origin of the RNG key. For scenario A and C, additional time for reaching the QSDN-Controller for receiving the routing vector is measured. However, this behavior is only visible if the waiting time for a key is shorter than the propagation delay, i.e. above the cut-off points. 

The proposed architecture has been shown to be feasible for a 20-node network, with an edge-positioned control node using a 1:1 packet-to-key ratio for encrypting the \ac{cm} traffic. The used routing protocol was found to be a key performance factor. A strong dependency on the key consumption of the deployed routing protocol was observed, with the different cut-off points between the reactive and proactive protocols differing by a factor of nearly seven ($340 \text{kps}/50 \text{kps}$). The investigated parameters did not reveal a major performance degradation that could be attributed solely to the proposed architecture, as the performance pressure is still mainly determined by the routing protocol.

\subsection{Takeaways and Recommendations}\label{chapter:takeaways}

\begin{table*}[ht]
    \centering
    \begin{tabular}{ L{3.7cm} | C{3.7cm} | C{3.7cm} | C{3.7cm} |}
        
        & Separately-protected architecture & 
        CM-as-a-service \newline architecture &
        CM-via-KMS  \newline architecture \\
        
        \hline
        \hline

        security of the \ac{cm} traffic enabled by & PQC/PSK & QKD & QKD \\
         \hline

        authentication of the \ac{cm} traffic given by  & respective certificates & respective certificates & already existent \\
         \hline
         
        supported key distribution scheme & forwarding \& centralized & forwarding \& centralized &  only forwarding \\
         \hline

        metadata gathering by message interception &  easy & easy & hard \\
         \hline
        
        robustness against \ac{dos} attacks depends on & separate network topology & separate network topology and QKD/KM network topology & QKD/KM network topology \\
         \hline
        
        robustness against protocols relaying on network flooding & high & high & weak \\
         \hline
    
        QKD key consumption & none & dependent on packet-to-key ratio & dependent on packet-to-key ratio \\
         \hline
                
        additional processing pressure on KM layer caused by the CM traffic & none & dependent on key consumption & two times as much as the \ac{cms} for packet-to-key ratios larger than 1:1 (e.g. OTP); else identical \\
         \hline
         \hline
        
        use case (recommended) & research or very low key generation rates &  when using a centralized key distribution scheme & high security scenarios  \\
         \hline

    \end{tabular}
    \caption{High level overview depicting properties of different \ac{cm} architectures leading to use cases when different architectures show superiority.}
    \label{tab:matrix_recommend}
\end{table*}

In Table \ref{tab:matrix_recommend}, a high-level overview of the analyzed architectures is given, leading to recommendations when an architecture shows an advantage.
The \ac{sp} architectures performs best in the \ac{km} layer as it does not use QKD keys to encrypt the \ac{cm} traffic. As the physical architecture closely matches the logical architecture, it is very robust for routing algorithms relaying on network flooding and \ac{dos} attacks. Based on this property, this architecture is also advantageous for centralized key distribution schemes \cite{ITU3803}, as dedicated connections to a central instance are already present, and metadata is inherently revealed by the protocol. In these scenarios, the controller node also functions as the central \ac{kms}, collecting intermediate \ac{km} transports. However, secure communication with the QSDN-Controller is difficult to implement, and significant metadata is exposed to adversaries. Furthermore, it features the property of a highly exposed central instance. Therefore, the \ac{sp} architecture is suitable for demonstrating QKDN functionality (e.g., research) or for scenarios with very low key generation rates.

In the \ac{cms} architecture, the communication of the \ac{cm} layer is secured by attaching the QSDN-Controller and the \ac{sdn}-Agent as SAEs. However, this approach still exposes the same metadata as the \ac{sp} architecture. Since communication occurs via two network layers, it leverages redundancy for robustness against \ac{dos} attacks. In terms of performance, it is dependent of the packet-to-key ratio. The performance of the scheme converges to that of the proposed architecture for low ratios (identical for one key per packet) and to that of the \ac{sp} architecture for key ratios lower or identical as used to secure the traffic in the application layer. This architecture can be advantageous in scenarios where a centralized key distribution system is used, when the metadata discussed is not considered sensitive and can be revealed. 
In both, the \ac{sp}- and \ac{cms} architecture, the processing pressure in the application layer can be alleviated by routing the \ac{cm} traffic through a dedicated management network. Note, that in both architectures, the \ac{sdn}-Agent can be realized by a network encryptor, i.e. it simply forwards the collected \ac{cm} traffic encrypted to the QSDN-Controller.
Implementing the \ac{cm} layer for key relaying to a QSDN-Controller offers several advantages for high-security \ac{qkdn} deployments. These include fewer interfaces to other layers, built-in authentication of \ac{cm} traffic, and minimal information exposure during eavesdropping (limited to predecessor and successor nodes). However, this comes at the cost of increased processing pressure. The feasibility of the architecture depends on the key generation rate and the routing protocol's key consumption. A challenge arises with centralized key distribution, as nodes must exchange node-specific keys with a central instance, eliminating the benefit of metadata hiding. 

Given the high-security and confidentiality requirements of \ac{qkdn} deployments, the proposed architecture is recommended when high-security requirement need to be fulfilled and enough keys are available, as it offers a significant increase in security without affecting the latency of the user's message. However, additional measures as attaching the QSDN-Controller to more than one \ac{km} (gateway-) node and positioning it in the topological center of the \ac{qkdn} as well as a proper routing protocol might to be taken to enhance its scalability, which scales with the ongoing research in enhancing the scalability of the \ac{km} layer.

\section{Summary and Conclusion}\label{chapter:summary}
In this work, we investigated different \acf{cm} architectures in \ac{sdn}-enabled \acp{qkdn}, focusing on their security and performance. We started by conducting an in-depth analysis of the routing characteristics of \ac{qkdn} which need to be considered in the design and evaluation of \ac{cm} architectures. These characteristics are shaped by the employment of trusted nodes and the concurrent use of distinct transmission modalities, specifically the quantum- and the classical channel, for the purpose of generating and exchanging secure keys. Our analysis showed that routing protocols and architectures utilized in classical telecommunication networks are not directly applicable to \acp{qkdn}, and that the \ac{km} layer in \acp{qkdn} is particularly vulnerable to attacks, making it a favorable target. The following section presents a comparative analysis of the advantages and disadvantages of two extant architectures and our proposed architecture, which we term \ac{cm}-via-KMS. The three architectures under consideration are: the \ac{sp} architecture, which employs a key generation mechanism distinct from \ac{qkd} to secure \ac{cm} traffic; the \ac{cms} architecture, which utilizes the key distribution service of the \ac{qkdn}; and our novel \ac{cm}-via-KMS architecture, which exploits the \ac{km} layer to enable secure exchange of \ac{cm} traffic. Using an existent \ac{cm} architecture serving as performance-baseline, we demonstrated the feasibility of the proposed architecture through a simulation. Together with previous analysis, we provide recommendations based on use cases for which the different architectures show superiority. These are for research- or very low key generation rates scenarios the \ac{sp} architecture, for \ac{qkdn} deploying a centralized key distribution scheme the \ac{cms} architecture and for high-security scenarios the new, \ac{cm}-via-\ac{kms}, architecture. Future work will entail a thorough investigation of the scalability characteristics of the proposed architecture including improvements, as well as the development and evaluation of routing algorithms that can be effectively integrated within its structure. Additionally, a hardware implementation of the proposed architecture will be established to validate its practicality and performance.

\section*{Funding}
This project has received funding from the German research ministry "Bundesministerium fuer Bildung, Wissenschaft, Forschung und Technologie" (BMBF) as part of the DemoQuanDT research and innovation programm under grand agreement No. 16KISQ074.

\section*{Acknowledgement}
We would like to thank Stefan Röhrich from Rohde \& Schwarz Cybersecurity GmbH, for the useful discussions and comments that greatly improved the manuscript. 

\section*{Data availability statement}
Data underlying the results presented in this paper are not publicly available at this time. However, if there is sufficient interest, we plan to make a public version available.

\section{Appendix A: Simulation Framework}\label{chapter:sim_framework}

Considerable research efforts are being devoted to the development of software tools for simulating Quantum Key Distribution Networks (QKDNs). The modeling of quantum channels encompasses a wide range of approaches, from highly abstracted representations, such as exponential decay models (e.g. $e^{\alpha \cdot l}$), to more complex loss models implemented in established libraries like ProjectQ \cite{ProjectQ} or NetSquid \cite{Coopmans}. In addition to frameworks focused on accurately simulating quantum communication, such as those mentioned above, there are also platforms that emphasize the engineering aspects of QKDNs, exemplified by QKDNetSim \cite{QKDNetSim}. For a comprehensive overview of the existing QKDN simulation platforms, we refer the reader to \cite{survey_qkd_net_sim}, which provides a detailed overview of the current state of the art.

\begin{figure}[htp]
    \centering
    \includegraphics[width=0.5\textwidth]{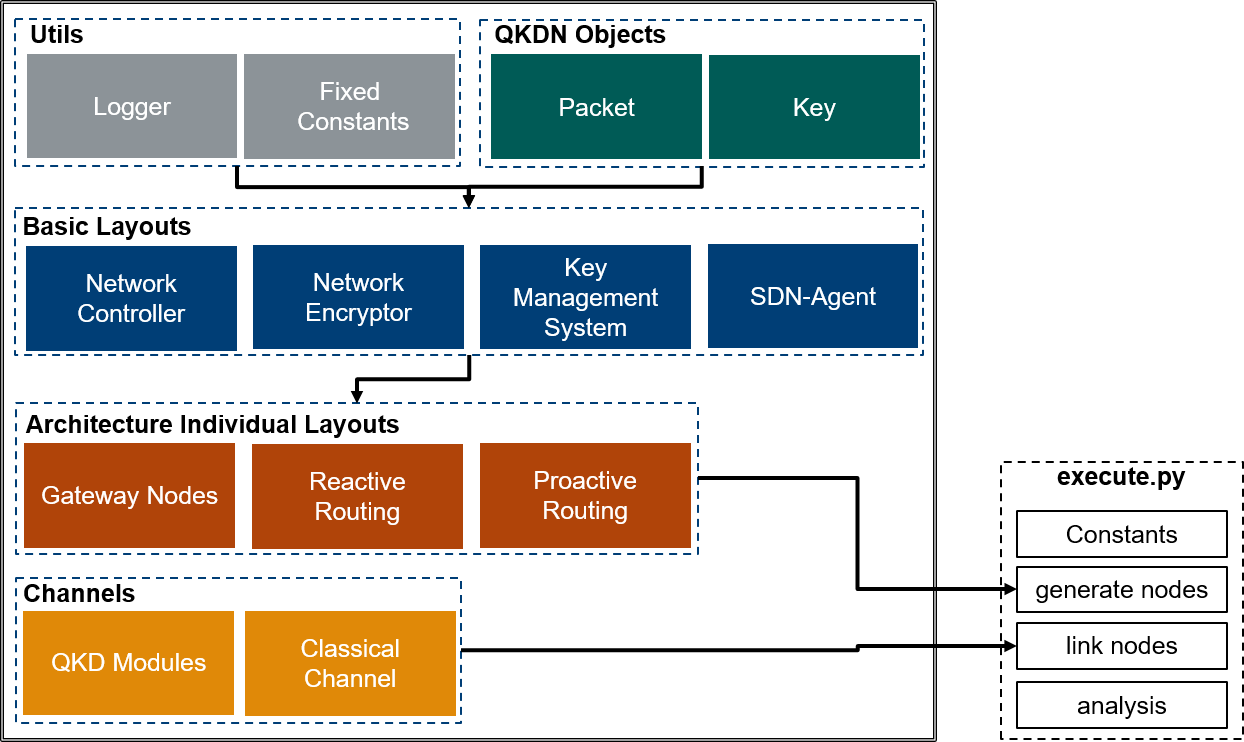}
    \caption{Structure of the presented simulator depicting its modular approach.
    To perform a simulation, the process involves the following steps: (1) defining a set of constants, which may include default values from the fixed constants in the utils section; (2) generating nodes using the relevant modules based on the architecture being investigated; (3) establishing links between the modules; (4) executing the simulation; and (5) logging and evaluating the results during and after execution, respectively.}
    \label{fig:layout_q_sec_net_sim}
\end{figure}

However, the existing simulators do not incorporate the architectures and parameters of Rohde \& Schwarz devices.  Moreover, the current frameworks are not well-suited for rapidly exploring new concepts and architectures due to their complex, underlying infrastructure. To address this limitation, we have developed a custom discrete-event network simulator using the Python framework \textit{SimPy}. This tool enables rapid prototyping and evaluation of different control- and management Layer implementations, and facilitates the development of new architectures and protocols. The simulator is characterized by the fact that it contains the architectures and parameters of the Rohde \& Schwarz off-the-shelf device SITLine ETH Classic, which makes it unique \cite{network_enc_rohde_schwarz}.  In the following the simulation framework is presented, cf. Figure \ref{fig:layout_q_sec_net_sim}.
We begin by outlining the fundamentals and characteristics, followed by an explanation of the various modules, and finally conclude with a validation of the simulation framework using the well established QKDNetSim tool.

\subsection{Framework}
The setup is distinguished by a modular framework that can be flexibly configured to accommodate a wide range of customized scenarios. Every single node may contain at least one QKD-Module, a Key-Management-System (KMS), an interface to the control- and management plane and potentially a cryptographic application (here, we currently only support the network encryptor (NE) SITLine Classic). Together with a network controller, the individual simulation framework is realized. In this framework, the modules are abstracted according to their phenomenological description and by parameters that are tailored to the user's needs. The phenomenological description primarily focuses on timeouts, e.g. for packet encryption and transmission times, which makes the discrete-event simulation framework \textit{SimPy} a suitable choice. Furthermore, it enhances the reproducibility of the research, enabling validation and extension of the findings. As the primary objective of this framework is to efficiently analyze various CM layer implementations, this level of detail is sufficient.

\subsection{QKD-Module}
The focus of the simulation is on the "key usage" aspects in a QKDN, so the QKD-Modules are abstracted to their key streaming behavior. A random string of zeros and ones is generated at a specified frequency, with the length of the string varying randomly around a given value (jitter), for accounting environment and protocol specific variations. 
Post-processing, simulated as a timeout, occurs when there are sufficiently enough bits. For a length matching the frequency, this value can be tuned to the unit keys per second, which is used in the simulation experiment Section \ref{chapter:simulation}. The resulting "secure" string is chopped into key-bits of a user-defined size and pushed towards the KMS. Keys exceeding the storage size of the KMS are ignored.
To prevent asymmetric key usage, the abstracted QKD-Modules split the generated keys according to their ID into even and uneven. The two parts are now pushed towards an interface reserved for keys at the KMS and used for either encryption or decryption \cite{me_etsi_sin}.

\subsection{Key Management System}
The KMS waits for the keys and triggers a setup message (sent to the QSDN-Controller), when a certain amount of keys is available (default=1). Upon request from its connected NE, a key exchange is performed to achieve a secure connection with the requested partner. The requests are realized as packets which are exchanged in the manner of the forwarding key distribution scheme using a random-number-generator (RNG) \cite{ITU3803}. We selected this scheme as it includes the flexibility to encrypt multiple keys separately or bundled using one QKD key, as well as the possibility to change the key size very easily which can be used for adapting key request bursts and as well as for reducing the network load on the \ac{km} layer.

\subsection{Network Encryptor}
The NE is requesting keys from the underlying KMS for sending encrypted messages to other NEs within the network. This is achieved by sending a request message to the connected KMS module, triggering a key transport via the network to the requested node. After a successful key transport, i.e. the relayed key is acknowledged, the NEs receive the keys from the connected KMS modules. Secure communication is now possible between the two NEs.
%

\subsection{Network Controller}
This single-existent instance is controlling the network as well as the simulation. Initially, it distributes routing information (e.g. routing tables, as well as translation tables between the KMS instances and their connected NEs). It also provides routing vectors on demand for source-reactive routing protocols and periodically updates routing tables for distributed-proactive routing protocols. The controller facilitates simulation monitoring by distributing logging files and tracking the status of all devices throughout the simulation. Once the maximum simulation time is reached, it stores the accumulated network information in a file, which is then forwarded for analysis.

\subsection{Classical Channel}
To interconnect distant devices, an instance, namely \textit{Classical Channel}, transmits the network packets from one device to another, while devices located at the same node/access point are assumed to communicate directly. A probability for dropping a package as well as a delay for realizing the distance can be set. For accounting the properties of a layer 2 NE, the classical channel instance collects the packets from the NEs output and forwards it according to an internal routing table received by the network controller, such that the sending NE does not need to care on proper routing. In order to avoid packet collisions, an exponential distributed backoff time (default=3s) is introduced, which may cause statistical variations in the data measurements. For example, a large backoff time causes no packet collisions, however it may influence the packet latency. In contrast to a backoff time chosen too low, causing an infinite loop as parallel arriving packets need to be re-transmitted. Statistical influence of the recorded measurements can thus be not fully eliminated, as the initialization seed of a distribution can cause random sequences which influence the average measurement result, e.g. by multiple nodes requesting a key from the same node. Simulation results are obtained by running the simulation multiple times with different random seeds and calculating the average value. 

The modular approach provides valuable insights in the design of QKDN, as well as the internal processing of the different network units, so that a variety of parameters can be captured during a simulation. Through the development of this simulator, we are investigating the design process of QKDN by gaining insight into the internal processing of the various devices. Furthermore, it allows for rapid prototyping of architectures and protocols, taking into account the internal architectures and parameters of Rohde \& Schwarz SITLine devices \cite{network_enc_rohde_schwarz}. A limitation is given by the computational efficiency of the simulator as well as the abstraction level.

\subsection{Validation}
To validate the simulation tool presented, we replicated the Padua Network using the simulation parameters from QKDNetSim \cite{QKDNetSim}. QKDNetSim was choosen as reference due to its its well-established status, its emphasis on engineering aspects, and its reported accuracy. The validation is carried out in a step-by-step manner, beginning with a comparison of the environment, followed by a comparison of the NEs in the application layer and the QKD-Modules in the quantum layer. In a last step, the KMS in the KM layer located in between is analyzed and compared.

\begin{figure}[btp]
    \centering
    \includegraphics[width=0.5\textwidth]{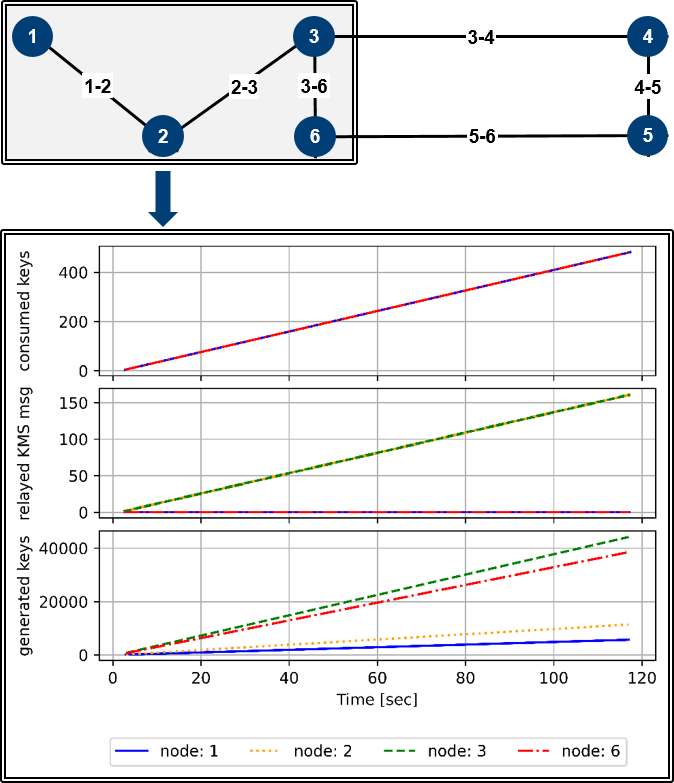}
    \caption{Schematic overview of the KM layer of the simulated Padua Network. It depicts the part of the Padua Network topology, used to establish an AES-256 secured connection between nodes 1 and 6, consists of two lower-key-rate free-space optical links (links 1-2 and 2-3) and a higher-key-rate fiber channel (link 3-6). The three graphs depict the amount of consumed keys, relayed KMS messages and generated keys.}
    \label{fig:schematic_padua}
\end{figure}

As the simulation framework only supports the devices from Rohde \& Schwarz, One-Time-Pad (OTP) Encryption is not supported. Therefore we restrict the validation to the application linking node 1 and node 6 (i.e. link '1-6') which uses the AES-256 encryption standard, cf. Figure \ref{fig:schematic_padua} \cite{QKDNetSim}. Therefore, the network effectively forms a path topology consisting of four nodes. For receiving the routing information, the \ac{sp} architecture using a distributed-reactive routing protocol was used. By deploying the simulation framework and configuring the devices with the reported parameters, we achieve a close match with the parameters reported by QKDNetSim.

\begin{table}[htp]
    \centering
    \begin{tabular}{| c | c | c | c |} 
        \hline
           & consumed keys & TX packets & RX packets  \\
        \hline
         QKDNetSim & 480 & 179.703 & 179.699 \\
         this work & 482 & 179.998 & 179.989 \\
         \hline
    \end{tabular}
    \caption{A comparison of the simulation results for the cryptographic application running on node 1 for QKDNetSim and the proposed simulation framework.}
    \label{tab:sim_compare}
\end{table}

Similar to QKDNetSim \cite{QKDNetSim}, the total runtime consists of a setup time followed by an effective runtime. Based on the provided data rate ($10,000$ kbps), packet size ($800 \cdot 8$ bits), and number of sent packets ($179.703$ packets), we can infer an effective runtime of $115$ seconds. From the in QKDNetSim reported total runtime of $120$ seconds, we can infer a setup time of $5$ seconds, during which a virtual connection between the nodes is established. Our setup phase is composed of the time to generate at least one key, the negotiation of the master-/slave roles and the as well as distribution of routing information. The node initialization in the setup phase includes the setup of key storages. Here, we report a (measured) setup time of roughly $2.5$ sec. we estimate the total runtime to be approximately 117.5 seconds, based on the effective runtime of 115 seconds. In contrast to QKDNetSim, our simulation framework initiates message transmission and reception after the setup phase, which means we cannot account the reported missed calls during this phase \cite{QKDNetSim}.
The values in Table \ref{tab:sim_compare} describe the behaviour of a NE running in the simulated Padua Network. As only one connection is existent, the NE at node 6 reports the identical parameters.
Comparing the total number of sent (TX) and received (RX) packets, between both simulation frameworks a high accuracy is identified. Only a comparably small difference of only $295$/$290$ packets for the TX/RX packets is measured, which can be attributed to the intrinsic differences of both frameworks as well as differences in the setup phase. Discrepancies between sent and received packets in our simulation framework can be attributed to encryption and propagation delays, which in turn impact the key consumption. Specifically, an additional key is required due to the packet-to-key ratio of 375:1, while another key is accounted for by the inherent behavior of the simulated device, which consistently maintains an extra key, resulting in a total of $482$ keys. Following the validation of applications requesting secure key material, we continue with the comparison of the generation of secure key material by the QKD-Modules. 
In QKDNetSim, the key generation process is modeled by mimicking the traffic from existing post-processing systems. This advanced QKD-Module abstraction precludes a direct comparison of the generated key material. However, we calculated from the total generated keys the respective key generation rate to $\frac{39023 \text{keys}}{115 \text{sec}}$ and $\frac{5820 \text{keys}}{115 \text{sec}}$ and were able to generate close results, cf. Table \ref{tab:verify_link_stats}. QKDNetSim generated bits corresponding to $5.820$ keys for links 1-2 and 2-3, and $39.023$ keys for link 3-6, each with a 256-bit length. In contrast, our measurements yield $5.684$ keys (97.7\%) and $38.512$ (98.1\%) keys, respectively. Compare Figure \ref{fig:schematic_padua} for the amount of consumed keys, relayed KMS messages and generated keys over the simulation time.

\begin{table}[htp]
    \centering
    \begin{tabular}{| c | c | c | c |}
    \hline
        Link & key rate (keys/sec) & generated keys & relayed keys \\
        \hline
         1-2 & 58 & 5684 & 483 \\ 
         2-3 & 58 & 5684 & 483 \\ 
         3-6 & 390 & 38.512 & 483 \\
         \hline
    \end{tabular}
    \caption{Link Statistics: dependent on the link, a different key rate is set mimicking the results of the generated key bits reported by QKDNetSim \cite{QKDNetSim}. The amount of relayed keys show the number of keys requested for supplying the cryptographic application between node 1 and node 6. The generated keys exhibit a high degree of accuracy, matching the reported keys with high precision.}    
    \label{tab:verify_link_stats}
\end{table}

In a last step, we compare the behaviour of the KMS. Differences in the amount of relayed keys can be attributed to the amount of keys stored in the NE (here 3), the threshold when new keys are requested (here 0) as well as the number of keys per request (here 3). QKDNetSim here reports $290$ keys à $512$bit (i.e. $580$ keys à $256$ bit) which are transformed locally on the needed size, while our NE requested $483$ keys à $256$bit \cite{QKDNetSim}. 
According to the results, we can conclude that the developed simulation tool yields reliable results when compared to the QKDNetSim simulation framework which is sufficient for our purposes.

\section{Appendix B: Simulation Assumptions and Parameters}\label{chapter:appendix_sim_body}

\subsection{Assumptions}
For investigating the feasibility of the proposed architecture, the simulation deploys 20 nodes arranged in a topology resembling the network properties of the internet \cite{internet_like_topology}. The following assumptions and parameters are selected such that the bare architecture, i.e. without additional measures for enhancing its scalability (c.f. Chapter \ref{chapter:takeaways}), is stressed to the maximum in order to gain deeper insights into its performance without distorting the overall measurement result. The simulation is carried out by a \ac{qkdn} simulator in an Windows 10 environment on a Intel i7 chip, cf. Appendix \hyperref[chapter:sim_framework]{A}. An overview of the parameters used can be found in Table \ref{tab:param_network}. Further assumptions include,

\begin{itemize}
    \item no changes/updates in the routing tables/vectors according to the actual node status, i.e. to only focus on the bare protocol and not being algorithm specific

    \item no delay influence from the management network in the \ac{sp} architecture on the \ac{qkdn}, i.e. no delay caused by message encryption
    
    \item neglecting the possible packet drop rate
    
    \item only implementing the routing protocol in the \ac{km} layer, by assuming Layer 2 NEs in the application layer
    
    \item an edge position for the controller node (cf. Figure \ref{fig:network_topo}), to stress the performance pressure on the gateway node
    
    \item the distances between all connected nodes are identical, i.e. identical propagation delays
    
    \item continuous bit rate, where messages arrive at a constant rate, resulting (nearly) in periodic key requests 
    
    \item eleven source/access nodes and nine relaying-only (backbone) nodes
    
    \item no delay caused by the controller as it is designed for such network loads

    \item a 1:1 packet-to-key ratio for the \ac{cm} traffic for gaining deeper insights into the new architecture 
    
\end{itemize}

\begin{table}[htp]
\centering
    \begin{tabular}{|l|c|c|}
    \hline
    \textbf{Variable}        & \textbf{Unit} &  \textbf{Value} \\
    \hline
    \textit{QKD-Module}        &  &   \\
            key generation rate jitter   & \%   & 5  \\
            post processing duration   & sec   & 1  \\
    \hline 
    \textit{KMS-Module}        &  &   \\
            storage size & keys & $10^5$  \\
            encryption latency & ms & 0.018 \\
            link status updates & sec & 60 \\
            key size & bits & 256 \\ 
    \hline
    \textit{Network Encryptor}        &  &   \\
            encryption latency per packet & ms & 0.018 \\
            arrival rate & kbits & 10 000 \\
            packet size & byte & 800 \\
            key life time & sec & 0.24 \\
            number of keys per request & - & 3 \\
    \hline
    \textit{Classical Channel}        &  &   \\
            delay & ms  & 2  \\
            loss probability & $\%$  & 0  \\
    \hline
    \textit{Runtime}        &  &   \\
            total simulation time & sec  & 400 \\
    \hline
\end{tabular}
     \caption{Main Parameters used for the simulation. The parameters were selected close or identical to the ones given by the Padua Network (given in \cite{QKDNetSim}). }
    \label{tab:param_network}
\end{table}

\subsection{Topology}

\begin{figure}[htp]
    \centering
    \includegraphics[width=0.5\textwidth]{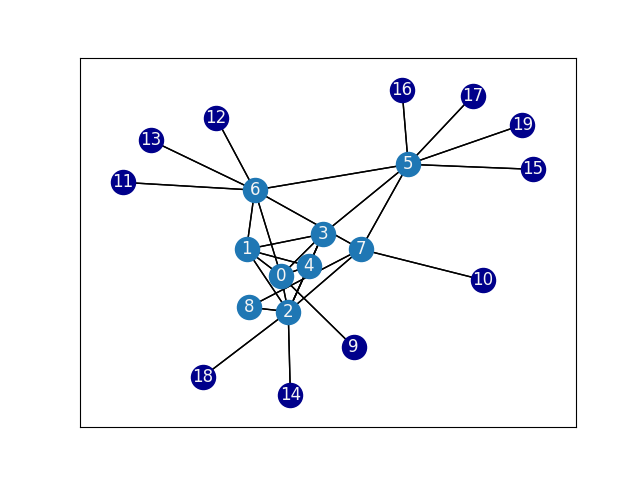}
    \caption{The investigated network topology consists of 20 nodes, designed to mimic Internet properties (seed=42) \cite{internet_like_topology}. Dark blue nodes are access/source nodes that send and receive messages, while light blue nodes are relaying-only backbone nodes. For scenarios C and D, a gateway node is added next to Node 19 (adjacent to Node 5) to facilitate access to the QSDN-Controller.}
    \label{fig:network_topo}
\end{figure}

Furthermore, the identical network topology was used for both the application and \ac{km} layers, as shown in Figure \ref{fig:network_topo}. A star topology for the management traffic was deployed, with the QSDN-Controller at its center.

\begin{thebibliography}{10}
\newcommand{\enquote}[1]{``#1''}
\providecommand\JournalTitle[1]{#1}

\bibitem{Mehic}
M.~Mehic, M.~Niemiec, S.~Rass, J.~Ma, M.~Peev, A.~Aguado, V.~Martin, S.~Schauer, A.~Poppe, C.~Pacher, and M.~Voznak, \enquote{Quantum key distribution: A networking perspective,} {\protect\JournalTitle{ACM Comput. Surv.}} \textbf{53} (2020).

\bibitem{ITU3803}
ITU-T, \enquote{ITU-T Y.3803 Quantum Key Distribution Networks – Key Management,} Tech. rep., ITU-T (2021).

\bibitem{wdm_routing}
Y.~Cao, Y.~Zhao, C.~Colman-Meixner, X.~Yu, and J.~Zhang, \enquote{Key on demand (kod) for software-defined optical networks secured by quantum key distribution (qkd),} {\protect\JournalTitle{Opt. Express}} \textbf{25}, 26453--26467 (2017).

\bibitem{Sharma}
P.~Sharma, V.~Bhatia, and S.~Prakash, \enquote{Priority order-based key distribution in qkd-secured optical networks,} in \emph{2020 IEEE International Conference on Advanced Networks and Telecommunications Systems (ANTS),}  (IEEE Press, 2020), p. 1–6.

\bibitem{stoachstic_routing}
M.~Li, D.~Quan, and C.~Zhu, \enquote{Stochastic routing in quantum cryptography communication network based on cognitive resources,} in \emph{2016 8th International Conference on Wireless Communications \& Signal Processing (WCSP),}  (2016), pp. 1--4.

\bibitem{ada_china_routing}
L.-Q. Chen, M.-N. Zhao, K.-L. Yu, T.-Y. Tu, Y.-L. Zhao, and Y.-C. Wang, \enquote{Ada-qkdn: a new quantum key distribution network routing scheme based on application demand adaptation,} {\protect\JournalTitle{Quantum Information Processing}} \textbf{20} (2021).

\bibitem{qos_routing}
M.~Mehic, P.~Fazio, S.~Rass, O.~Maurhart, M.~Peev, A.~Poppe, J.~Rozhon, M.~Niemiec, and M.~Voznak, \enquote{A novel approach to quality-of-service provisioning in trusted relay quantum key distribution networks,} {\protect\JournalTitle{IEEE/ACM Transactions on Networking}} \textbf{28}, 168--181 (2020).

\bibitem{sdn_survey}
D.~Kreutz, F.~M.~V. Ramos, P.~E. Veríssimo, C.~E. Rothenberg, S.~Azodolmolky, and S.~Uhlig, \enquote{Software-defined networking: A comprehensive survey,} {\protect\JournalTitle{Proceedings of the IEEE}} \textbf{103}, 14--76 (2015).

\bibitem{sdn_qkd_trend1}
A.~Aguado, V.~Lopez, J.~Martinez-Mateo, T.~Szyrkowiec, A.~Autenrieth, M.~Peev, D.~Lopez, and V.~Martin, \enquote{Hybrid conventional and quantum security for software defined and virtualized networks,} {\protect\JournalTitle{J. Opt. Commun. Netw.}} \textbf{9}, 819--825 (2017).

\bibitem{sdn_qkd_trend2}
A.~Aguado, V.~Lopez, J.~Martinez-Mateo, M.~Peev, D.~Lopez, and V.~Martin, \enquote{Virtual network function deployment and service automation to provide end-to-end quantum encryption,} {\protect\JournalTitle{J. Opt. Commun. Netw.}} \textbf{10}, 421--430 (2018).

\bibitem{ITU3805}
ITU-T, \enquote{ITU-T Y.3805 Quantum Key Distribution Networks – Software-defined Networking Control,} Tech. rep., ITU-T (2021).

\bibitem{etsi_015}
E.~G.~Q. 015, \enquote{QKD Control Interface for Software defined Networks,} Tech. rep., see portal.etsi.org (2022).

\bibitem{madrid_qkdn}
M.~Irene, G.~Cid, and L.~O. Mart{\'{i}}n, \enquote{{MADRID QUANTUM NETWORK : A first step to quantum internet},} {\protect\JournalTitle{ACM}}  (2021).

\bibitem{sdn_uk}
R.~S. Tessinari, R.~I. Woodward, and A.~J. Shields, \enquote{Software-defined quantum network using a qkd-secured sdn controller and encrypted messages,} in \emph{Optical Fiber Communication Conference (OFC) 2023,}  (Optica Publishing Group, 2023), p. W2A.38.

\bibitem{etsi_014}
E.~G.~Q. 014, \enquote{Quantum Key Distribution (QKD); Protocol and Data Format of REST-based Key Delivery API,} Tech. rep., see portal.etsi.org (2019).

\bibitem{survey_routing_protocols_2}
E.~Alotaibi and B.~Mukherjee, \enquote{A survey on routing algorithms for wireless ad-hoc and mesh networks,} {\protect\JournalTitle{Computer Networks}} \textbf{56}, 940--965 (2012).

\bibitem{ProjectQ}
D.~S. Steiger, T.~Häner, and M.~Troyer, \enquote{Projectq: an open source software framework for quantum computing,} {\protect\JournalTitle{Quantum}} \textbf{2}, 49 (2018).

\bibitem{Coopmans}
T.~Coopmans, R.~Knegjens, A.~Dahlberg, D.~Maier, L.~Nijsten, J.~D.~O. Filho, M.~Papendrecht, J.~Rabbie, F.~Rozp, M.~Skrzypczyk, L.~Wubben, W.~D. Jong, D.~Podareanu, A.~Torres-knoop, D.~Elkouss, and S.~Wehner, \enquote{{NetSquid , a NETwork Simulator for QUantum Information using Discrete events},} {\protect\JournalTitle{Nature Publishing Group}}  (2015).

\bibitem{QKDNetSim}
E.~Dervisevic, M.~Voznak, and M.~Mehic, \enquote{Large-scale quantum key distribution network simulator,} {\protect\JournalTitle{J. Opt. Commun. Netw.}} \textbf{16}, 449--462 (2024).

\bibitem{survey_qkd_net_sim}
A.~Aji, K.~Jain, and P.~Krishnan, \enquote{A survey of quantum key distribution (qkd) network simulation platforms,} in \emph{2021 2nd Global Conference for Advancement in Technology (GCAT),}  (2021), pp. 1--8.

\bibitem{network_enc_rohde_schwarz}
Rohde \& Schwarz Cybersecurity GmbH, \enquote{R\&S SITLine ETH-L,}  (2022).

\bibitem{me_etsi_sin}
P.~Horoschenkoff, J.~Rödiger, S.~Röhrich, and M.~Wilske, \enquote{On the security and performance assessment in the design of quantum key distribution networks,} in \emph{10th ETSI/IQC Quantum Safe Cryptography Conference,}  (2024).

\bibitem{internet_like_topology}
A.~Elmokashfi, A.~Kvalbein, and C.~Dovrolis, \enquote{On the scalability of bgp: The roles of topology growth and update rate-limiting,} in \emph{Proceedings of the 2008 ACM CoNEXT Conference,}  (Association for Computing Machinery, New York, NY, USA, 2008), CoNEXT '08.

\end{thebibliography}
\end{document}